\documentclass[noeprint,noshowpacs,nopreprintnumbers,twocolumn,prb,showpacs,superscriptaddress,amsmath,amssymb,citeautoscript,aps,10pt]{revtex4-1}
\usepackage{graphicx}
\usepackage{dcolumn}
\usepackage{color}
\usepackage{bm}
\usepackage{braket}
\usepackage[hidelinks]{hyperref}
\hypersetup{
    colorlinks,
    citecolor=blue,
    filecolor=blue,
    linkcolor=blue,
    urlcolor=blue
}
\usepackage{lipsum}

\allowdisplaybreaks

\begin{document}

\title{Non-adiabatic transformation of a spin-chain geometry via local control}

\author{P.~V.~Pyshkin}
\affiliation{Institute for Solid State Physics and Optics, Wigner Research Centre, Hungarian Academy of Sciences, P.O. Box 49, H-1525 Budapest, Hungary}
\author{E.~Ya.~Sherman}
\email{evgeny.sherman@ehu.es}
\affiliation{Department of Physical Chemistry, The University of the Basque Country UPV/EHU, 48080 Bilbao, Spain}
\affiliation{IKERBASQUE, Basque Foundation for Science, 48011 Bilbao, Spain}
\author{Lian-Ao Wu}
\affiliation{IKERBASQUE, Basque Foundation for Science, 48011 Bilbao, Spain}
\affiliation{Department of Theoretical Physics and History of Science, The University of the Basque Country UPV/EHU, 48080 Bilbao, Spain}

\date{\today}

\begin{abstract}
We consider transformation from a closed to an open spin chain and vice versa produced by changing single
link strength in a pair of neighboring spins. We show that in the non-adiabatic time domain fidelity of such a process can be increased 
 by proper choosing of the control function for spin-spin exchange coupling. 
We obtain this function for an antiferromagnetic quantum Ising chain and present heuristic reasons restricting possible 
time-dependences of Hamiltonians applied for a high-fidelity control.
\end{abstract}

\maketitle

\section{Introduction}

Precise control of complex quantum systems became important task in last decades due to the prospective 
of building a quantum computer~\cite{Nielsen2000} and experimental feasibility of such kind 
of control~\cite{Busl2013,Russ2017}. Entanglement is the main feature of quantum computing algorithms, 
thus producing and destroying entanglement between parts of a complex system~\cite{Pyshkin2018-NJP} 
is a problem of importance and interest. Another type of problems, related to manipulation of entanglement, 
is a transformation of a complex system from a highly entangled ground state of the initial
Hamiltonian~$H_i$ to a demanded state, which is the ground state of another
Hamiltonian~$H_f$. These Hamiltonians~$H_i$ and~$H_f$ can be related to different geometry 
(or topology) of a complex system. We would like to emphasize that although our goal is the transformation 
of {\em quantum states} one into another, we make a correspondence between initial and final  
states and {\em Hamiltonians}. For examples, investigating a spin lattice, we can consider it as a 
graph with spins as vertices, spin-spin interactions as edges, and in such a case~$H_i$ and~$H_f$ 
correspond to different configurations of the edges. 
If $H_i$~($H_f$) corresponds to a disconnected (connected) graph and $[H_i, H_f]\neq0,$ 
we have a modification in the system entanglement. 

The task of a quantum ground state transformation 
can be solved via unitary process using adiabatic control when evolution is governed by 
a time dependent Hamiltonian~$H(t) = a(t)H_i + b(t)H_f$, where $a(0)=1$, $a(T)=0$, $b(0)=0$, $b(T)=1$, and $T$ 
is an evolution time. This process is also the main recipe of adiabatic quantum computation \cite{Das2008, Adiabatic_qc_1}
and quantum annealing (see, e.g. [\onlinecite{Pino2018}] for quantum spin systems). 
In most cases it is sufficient to set linear switching: $a(t)=1 - t/T$ and $b(t) = t/T$ to get a demanded result. 
The problem here is that time~$T$ must be sufficiently long to satisfy the adiabatic theorem conditions~\cite{Born1928}. 
However, at a long evolution time the system undesired decoherence becomes possible, thus the challenge of adiabatic 
shortcuts appears. Some approaches to make adiabatic evolution shorter, such as 
counteradiabatic driving~\cite{Demirplak_adiab_drive, Berry-2009}, strength-pulsed (noise) 
control~\cite{Jing2014, HQC-Pyshkin2016} and others~\cite{Damski2014,Shortcuts-Torrontegui-Muga-2013}
have been presented recently.
A common feature of all these  techniques, which is difficult to realize,  
is the need of control all parts of the quantum system.
Note that in our task we do not need the quantum system to  
be in the instantaneous ground state of the intermediate Hamiltonian~$H(t)$ ($0<t<T$) during the 
entire transformation process since we are interested only in the final state. In present paper we consider 
transformation from the ground state of a closed spin chain to the ground state of an open spin chain
and show that it is possible to use local-only control of a special kind to increase
the fidelity of the target state in the non-adiabatic time domain. Some general restrictions imposed on the control
function will be presented and discussed.

\section{Results}
\subsection{The model} 
The key element of our investigation is a time-dependent antiferromagnetic quantum Ising Hamiltonian in a transverse field 
(with the spin-spin Ising coupling taken as the unit of energy):
\begin{equation}\label{Hamiltonian}
	H(g) = \sum_{n=1}^{N-1} \sigma_n^x\sigma_{n+1}^x + g\sigma_1^x\sigma_{N}^x + B\sum_{n=1}^N \sigma_n^z,
\end{equation}
where $\sigma^x,\, \sigma^z$ are the Pauli matrices, $B$ is an external magnetic field, $N$ 
is the number of spins in the chain, and $g$ is a time-dependent control parameter. 
We assume that quantum state of the chain is described by a vector~$\ket{\psi(t)}$, 
and the initial state is~$\ket{\psi(0)}=\ket{\phi_0}$, with $H(g=1)\ket{\phi_0}=\lambda_0\ket{\phi_0}$, and $\lambda_0$
being the minimum eigenvalue of~$H(g=1)$. 
Note that the system evolves in time due to nonzero value of the commutator: 
\begin{eqnarray}
&&\left[\vphantom{\frac{1}{1}}\sum_{n=1}^{N-1} \sigma_n^x\sigma_{n+1}^x + B\sum_{n=1}^N\sigma_n^z, \; g\sigma_1^x\sigma_{N}^x\right] \\
&&=2igB\left(\sigma_{1}^{y}\sigma_{N}^{x} + \sigma_{1}^{x}\sigma_{N}^{y} \right)\neq 0, \nonumber
\end{eqnarray}
and the ability to control the process depends on the evolution of the product of spin components at the edge
connecting the first and and the last spin.

\begin{figure}
	\includegraphics*[width=0.7\linewidth]{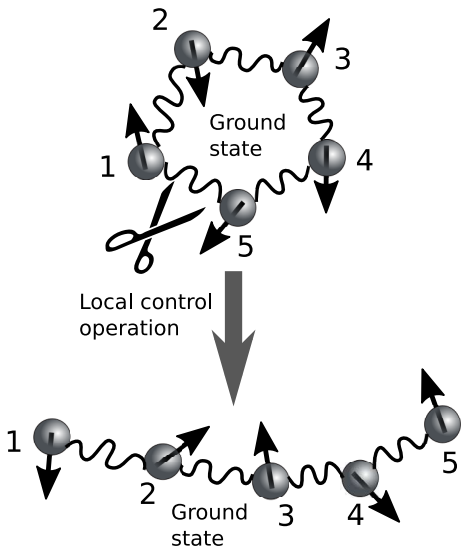}
	\caption{Schematic illustration of transformation from a closed to an open spin chain.}
	\label{fig_1}
\end{figure}

The problem now is to make a ``cut'' of a chain, 
i.e. to drive a system into a state $\ket{\psi(T)} = \ket{\chi_0}$, where $\ket{\chi_0}$ is the ground state 
of the open chain Hamiltonian~$H(g=0)$ (see Fig. \ref{fig_1}). We can write the final state as a result of unitary evolution
\begin{align}
	\ket{\psi(T)} = U(T) \ket{\psi(0)}, \\
	U(T) = \mathcal{T}\exp{\vphantom{1^1} \left( -i\int_{0}^{T}H(g(t'))dt'  \right) }, \label{t-exp}
\end{align}
where $\mathcal{T}$ stands for the time-ordering, and control function~$g(t)=1$ for $t\leq0$ and $g(t)=0$ for $t\geq T$.

We define the final target fidelity as follows
\begin{equation}
	f_{T} = |\braket{\chi_0|\psi(T)}|,
\end{equation}
and consider it as a functional of the control function:~$f_T = f_T[g(t)]$. The problem is 
to find the function~$g(t)$ maximizing the target fidelity~$f_T$ for fixed 
finite time~$T$. In order to obtain a proper control function we make a two-parametric 
parametrization at the interval $t\in[0,T]$:
\begin{equation}\label{g_parametrization_poly}
	g(t) = 1 - (1+a_1+a_2)\frac{t}{T} + a_1\frac{t^2}{T^2} + a_2\frac{t^3}{T^3},
\end{equation}
and the target fidelity becomes a function of control parameters~$f_T = f_T(a_1, a_2).$

\begin{figure}
	\includegraphics*[width=0.8\linewidth]{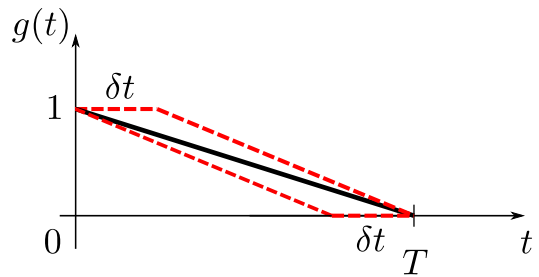}
	\caption{Schematic illustration of the fact that linear dependence is the optimal one for near-adiabatic regime.
	Red dashed lines correspond to the modifications of the linear decrease which is shown by a bold solid line. 
	Obviously, the modified control is not better for adiabaticity than the non-modified one.}
	\label{fig_2}
\end{figure} 

\subsection{The features of efficient control}
Before going further we emphasize that we restrict our consideration to the following condition: 
\begin{equation}\label{non_cross}
E_0(g)\neq E_1(g), \qquad \forall g \in (0,1),
\end{equation}
where $E_0(g)$ ($E_1(g)$) is the instant ground (first excited) state energy 
of the Hamiltonian~(\ref{Hamiltonian}). If the condition~(\ref{non_cross}) is valid then it is 
possible to make the adiabatic transformation with
\begin{equation}\label{adiab-limit}
	f_{T0}\equiv f_T(0,0)\rightarrow1, \quad T\rightarrow\infty.
\end{equation}
We can argue that for a large enough~$T$ a linear decrease~(\ref{adiab-limit}) is the optimal shape of the control. 
From the corresponding Schr\"{o}dinger equation in the adiabatic 
frame we can see that probability of leaving the instantaneous 
eigenstate of~$H(t)$ is proportional to the time derivative of 
Hamiltonian $|\partial H / \partial t | \propto |dg(t)/dt|$ (see e.g. textbook [\onlinecite{book_sakurai}]). 
Thus, the adiabaticity condition can be written as~$\partial H / \partial t\rightarrow0$, 
and if one changes~$g(t)$ for some fixed large~$T$ from linear dependence into 
some other shape it necessarily leads to increasing $|\partial H / \partial t|$ (for some $t\in(0,T)$), 
and consequently the adiabaticity condition tends to be broken. 
Another way to see that the linear decrease is the optimal control in the adiabatic domain is the following reasoning. 
Let us take a large enough~$T$ which corresponds to the adiabaticity regime for linear decrease, 
and periods~$T'$ with $T'>T$ also satisfy adiabaticity condition. 
Now, let us modify the control function a little in two different ways (see Fig. \ref{fig_2}):

\noindent 1) we start linear decrease not from $t=0$ but from $t=\delta t,$

\noindent or

\noindent 2) we finish linear decrease not at $t=T$ but at $t = T - \delta t.$

Both described modifications of the control function are nothing else than just decreasing 
the effective time~$T\rightarrow T-\delta t$. The latter means that adiabaticity with modified
control function will not be better than that for the linear shape, and by setting~$\delta t\rightarrow0$ 
we conclude that the linear decrease is the optimal shape for relatively large~$T$. This conclusion also 
can be seen from the numerical results in Ref.~[\onlinecite{Pyshkin2018-NJP}].

   \begin{figure}
   	\includegraphics*[width=0.7\linewidth]{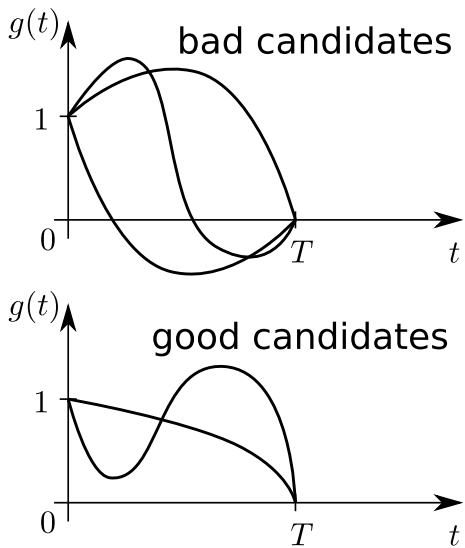}
   	\caption{Illustrative depiction of control functions 
   	not satisfying~(\ref{a_inequalities}) (upper panel) and satisfying~(\ref{a_inequalities}) (lower panel).}
   	\label{fig_3}
   \end{figure}

Our conclusion~(\ref{adiab-limit}) now will help us to understand the restrictions imposed on the control function~$g(t)$ in a non-adiabatic time domain.
We claim that any efficient control function must satisfy the following inequalities:
\begin{equation}\label{inequal_conditions}
	\left.\frac{dg(t)}{dt}\right\rvert_{t=0} < 0, \quad \left.\frac{dg(t)}{dt}\right\rvert_{t=T} < 0.
\end{equation}
The explanation of the restrictions~(\ref{inequal_conditions}) is the following. Let us assume that the optimal control 
function in the non-adiabatic domain does not satisfy~(\ref{inequal_conditions}).
Now we slightly increase the time~$T$ toward the adiabatic domain. As we have already noticed, the optimal~$g(t)$ shape 
in this region is a linear decrease~(\ref{adiab-limit}), thus, with the increasing~$T$ the optimal shape of~$g(t)$ must 
be continuously transformed into the linear shape. The latter means that there must be some intermediate values 
of~$T$ when $\left.dg(t)/dt\right\rvert_{t=0} = 0$ and $\left. dg(t)/dt\right\rvert_{t=T} = 0$. However, 
the control cannot be efficient if the derivative of the control function is zero at the start or at 
the end of a time interval, because one can consider that there is no control applied at all for  $t\ll T$ and $T - t \ll T$. 
Thus we obtain a contradiction with our initial statement 
that~(\ref{inequal_conditions}) is not valid for optimal control, and thus we conclude that~(\ref{inequal_conditions}) must be valid. 

By substituting~(\ref{g_parametrization_poly}) into the two-parametric Ansatz (\ref{inequal_conditions}) we get the following conditions:
\begin{equation}\label{a_inequalities}
	a_2 > -1 - a_1, \quad a_2 < \frac{1 - a_1}{2}.
\end{equation}
It is interesting to note that restrictions~(\ref{a_inequalities}) do not depend on time~$T,$ being in some sense 
the universal ones, and connect together adiabatic and
non-adiabatic time domains (even if we do not restrict the Ansatz in (\ref{g_parametrization_poly}) to cubic terms). 
Obviously, the point $(a_1=0, a_2=0)$ satisfies~(\ref{a_inequalities}). In Fig. \ref{fig_3} we show 
an illustrative example of control functions which are ``bad candidates'' (not satisfying~(\ref{a_inequalities}))  
and ``good candidates'' (satisfying~(\ref{a_inequalities})).

Although this analysis is not mathematically rigorous, it provides the solid basis for the understanding of the  
optimal control. Further numerical results presented below confirm our statement~(\ref{inequal_conditions}).

\subsection{Numerical results}
We use a numerical search in order to find the optimal values of~$a_1$ and~$a_2$ as
can be made by using two different approaches such as: 

\noindent i) by brute-force evaluation of~$f_T$ for all possible values of parameters (within some finite region and step size), 
in other words -- building a landscape of fidelity;

\noindent or 

\noindent ii) by using gradient search algorithms (for calculations in this paper we used the Broyden-Fletcher-Goldfarb-Shanno (BFGS) 
optimization method~\cite{BFGS} included in the Python scientific packages). 

Numerical calculations of time-ordered exponent~(\ref{t-exp}) have been done by dividing the 
time interval~$T$ to~$M\gg 1$ pieces~$\Delta t_{M} = T/M$, and using approximation
\begin{equation}\label{num_propagator}
U(T)\approx \prod_{n=0}^{M-1}\exp\left\{\vphantom{\frac{1}{1}}-i\Delta t_{M} 
H\left(\vphantom{1^1}g(\Delta t_{M}\left(n+{1}/{2})\right)\right)\right\}.
\end{equation}
Here we took~$M=300$ and matrix exponents have been calculated via a built-in Python function.

\begin{figure}
	\includegraphics*[width=0.8\linewidth]{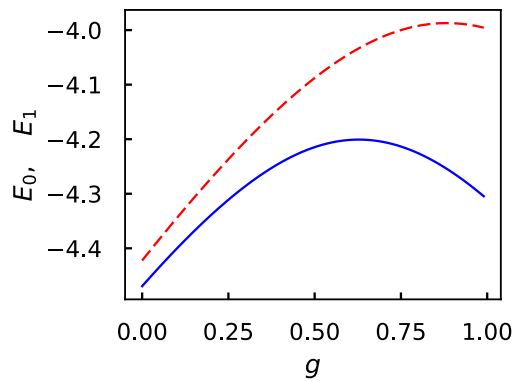}
	\caption{The ground ($E_{0},$ blue solid line) and the first excited ($E_{1},$ red dashed line) state energy of the Hamiltonian~(\ref{Hamiltonian}) 
	as a functions of~$g$ parameter. Here $N=5$ and~$B=0.5$.}
	\label{fig_4}
\end{figure}

Although the first approach needs large computational efforts, 
building the landscapes for searching of the optimized control functions can be done by parallel computing tools. 
In this paper we show that these two approaches give the same result. For numerical analysis we use the following Hamiltonian 
parameters: $N=5$ and $B=0.5$. With this choice we satisfy the level non-crossing condition~(\ref{non_cross}) 
and in Fig. \ref{fig_4} present the dependence of the two lowest eigenvalues of the Hamiltonian~(\ref{Hamiltonian}).

\begin{figure}
	\includegraphics*[width=0.8\linewidth]{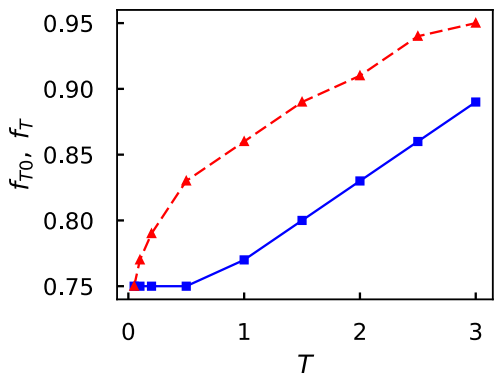}
	\caption{Output fidelity of non-optimized (blue solid line) and optimized (red dashed line) transitions 
	for cutting a spin chain with~$N=5$ and~$B=0.5$.}
	\label{fig_5}
\end{figure}

In Fig. \ref{fig_5} we present the non-optimized fidelity~$f_{T0}$ and the~$f_T$ obtained with 
optimized~$g(t)$ found by the gradient search. We see that advantage of the using of optimized 
control appears in a wide range of non-adiabatic~$T.$ 

\begin{figure}
	\includegraphics*[width=0.95\linewidth]{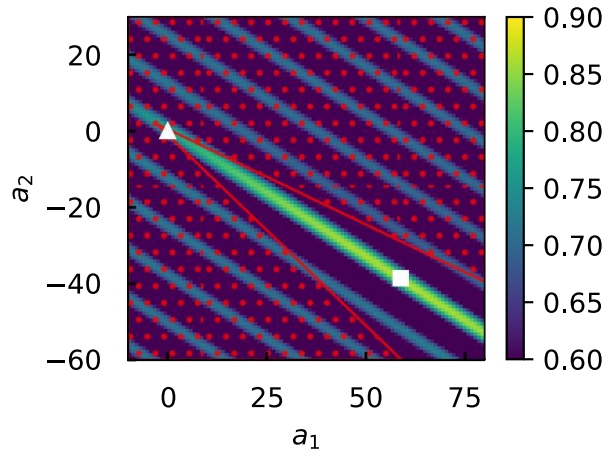}
	\caption{Landscape of the target fidelity $f_{T}(a_{1},a_{2})$ for~$T=1$. 
	White triangle corresponds to the simple linear control and square corresponds to the maximal fidelity process. Red solid lines correspond 
	to the boundaries in (\ref{a_inequalities}).}
	\label{fig_6}
\end{figure}

Let us look into a landscape of the fidelity in the parameter space for~$T=1$ as depicted in Fig. \ref{fig_6}. 
White triangle corresponds to a linear decrease and white square corresponds to an effective control shape 
for this particular~$T$. Area covered by the red dots corresponds to range of parameters which does 
not satisfy~(\ref{a_inequalities}). As can be seen in this Figure, there is a continuous ``island'' of a high fidelity 
connecting linear decrease (white triangle) and the optimal shape (white square). 
Thus, the numerical gradient search started from the linear shape is efficient in this situation. 

\begin{figure}
	\includegraphics*[width=0.95\linewidth]{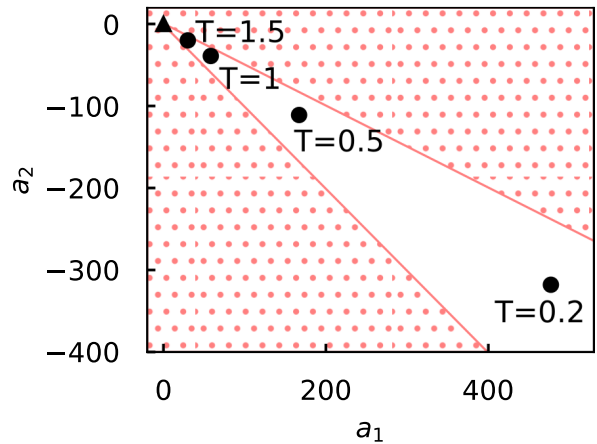}
	\caption{Optimal values of parameters~$a_1$ and~$a_2$ for different~$T$. 
	All of them lie in the region determined by inequalities~(\ref{a_inequalities}) (non-dotted area). 
	Triangle is the adiabatic limit of control with the linear decrease.}
	\label{fig_7}
\end{figure}

Now let us return to our statement~(\ref{adiab-limit}). Our numerical results confirm that the linear 
decrease is the optimal control in the adiabatic limit. In Fig.~\ref{fig_7} we show the dependence of 
optimal values of parameters~$a_1,\,a_2$ for different times~$T$. One can see that all points are
inside allowed by~(\ref{a_inequalities}) area, and they converge to the linear decrease (depicted as triangle). 

\begin{figure}
	\includegraphics*[width=0.95\linewidth]{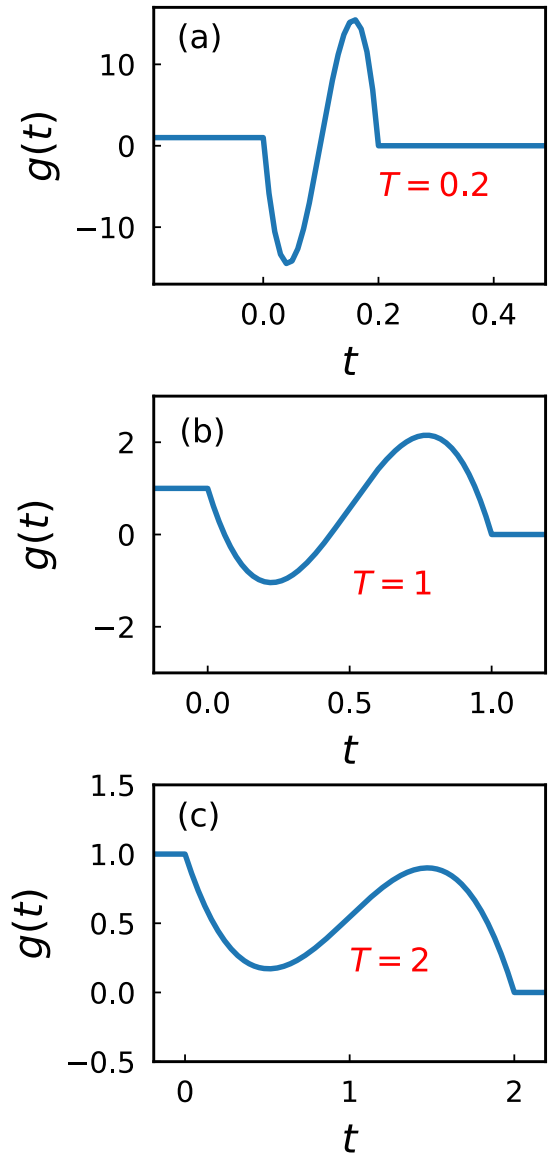}
	\caption{Optimal control shapes for different~$T$. All of them correspond to the ``good candidates'' in Fig. \ref{fig_3}.}
	\label{fig_8}
\end{figure}

In Fig.~\ref{fig_8} we show the optimal shapes of control for some values of time~$T$. 
As can be seen from Fig.~\ref{fig_8}, the calculated shapes correspond to a ``good candidates'' in~Fig. \ref{fig_3}.

\subsection{Pulsed control}

\begin{figure}
	\includegraphics*[width=0.95\linewidth]{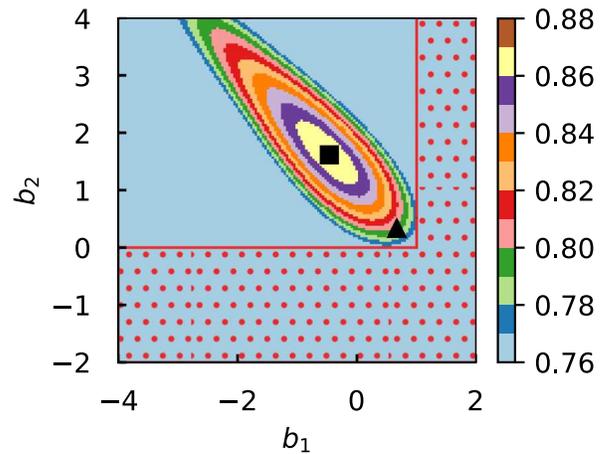}
	\caption{Landscape of the target fidelity in a parameter space for~$T=1$ for rectangular 
	pulsed control~(\ref{g-pulsed}) with $K=2$. Black triangle corresponds to a simple 
	``quasi-linear'' control ($b_1=2/3$ and $b_2=1/3$), and rectangular corresponds to the maximum output fidelity.}
	\label{fig_9}
\end{figure}

Inequalities~(\ref{inequal_conditions}) are written for a smooth control function. 
Now, let us assume that we apply for the control discontinuous rectangular pulses as:
\begin{equation}\label{g-pulsed}
	g(t) = \sum_{n=1}^K b_n
	\left[\theta \left(\vphantom{1^1}t - (n-1)\Delta t_{K}\right) - 
	\theta \left(t - n\Delta t_{K}\right)     \right],
\end{equation}
where we have~$K$ pulses with the duration~$\Delta t_{K} = T/K$, amplitude of $n$-th pulse is~$b_n$, 
and $\theta(t)$ is the Heaviside step function. In such a case by doing the same reasoning 
(for~$K\gg1$) as above we arrive at the following restrictions
\begin{equation}\label{b-inequality}
	b_1 < 1, \quad b_K > 0.
\end{equation}
However, we can see that restrictions~(\ref{b-inequality}) are valid even for small number of pulses.
In Fig. \ref{fig_9} we show the landscape of fidelity for pulse control~(\ref{g-pulsed}) for the case~$K=2$. 
Instead of comparing optimal parameters with the simple linear shape as
in~Fig. \ref{fig_6}, we assume ``quasi-linear'' step-like decrease with~$b_1=2/3$ and~$b_2=1/3$ as a 
non-optimized control function and as a starting point for the numerical gradient search.  

\subsection{Chain stitching with pulsed control}

We can investigate the opposite problem: stitching of a spin chain, i.e. transformation from the ground state 
of an open chain to the ground state of a closed one. In this case we can rewrite the 
restrictions~(\ref{inequal_conditions}) in the following form
\begin{equation}\label{stich_ineq}
		\left.\frac{dg(t)}{dt}\right\rvert_{t=0} > 0, \quad \left.\frac{dg(t)}{dt}\right\rvert_{t=T} > 0.
\end{equation}
For stitching under the pulsed control~(\ref{g-pulsed}) we have~$b_1>0$ and~$b_K<1$. 
In Table~\ref{t1} we put numerical results for cutting and stitching processes for 
pulse control with~$K=2$. As can be seen, the optimal values of~$b_1$ and~$b_2$ satisfy 
conditions (\ref{stich_ineq}) imposed on them.

\begin{table}[]
	\centering
	\caption{Non-optimized and optimized fidelity for pulse controlled cutting and stitching spin chain with $N=5$ and $B=0.5$.}
	\begin{tabular} {l|l|l|l|l}
		        & $T=1$ & $T=2$ & $T=3$ & $T=4$ \\ \hline 
		\multicolumn{5}{ c }{cutting the spin chain} \\ \hline
		$f_{T0}$& 0.80 & 0.88 & 0.95 & 0.99 \\
		$f_{T}$ & 0.87 & 0.92 & 0.97 & 0.99\\
		$b_{1}$ & -0.48 & 0.37 & 0.58 & 0.68 \\
		$b_{2}$ & 1.59 & 0.70 & 0.49 & 0.45 \\ \hline
		\multicolumn{5}{ c }{stitching the spin chain} \\ \hline
		$f_{T0}$& 0.80 & 0.88 & 0.95 & 0.99 \\
		$f_{T}$ & 0.87 & 0.92 & 0.97 & 0.99\\
		$b_{1}$ & 1.59 & 0.70 & 0.49 & 0.45 \\
		$b_{2}$ & -0.48 & 0.37 & 0.58 & 0.68		
	\end{tabular}
	\label{t1}
\end{table}
  
\section{Conclusions}

Despite the difficulties in analytical investigation of the time-ordered propagator~(\ref{t-exp}) 
we have found general restrictions~(\ref{inequal_conditions}) on the shape of efficient optimized 
control functions. These conditions follow from the assumption of existence of a continuous transformation 
from a non-adiabatic to the adiabatic time domain. This transformation is possible when the lowest energy levels of the 
Hamiltonian do not cross~(\ref{non_cross}). Thus, we connect the behavior of a complex quantum system 
in adiabatic and non-adiabatic time domains. Although our proposal is based on the locality of 
the applied control, our conclusions~(\ref{inequal_conditions}) are not limited either by the locality or by a special 
kind of a quantum system such as a spin chain. This analysis can be applied for speeding up transformation of 
any complex quantum system which can be transformed in the adiabatic way, 
including optical control of cold atoms in the quantum speed limit regime  \cite{Alberti2018}.


\section{Acknowledgement}

We gratefully acknowledge National Research, Development and Innovation Office of Hungary
(Project Nos. K124351 and 2017-1.2.1-NKP-2017-00001), the Basque Country Government (Grant No. IT472-10), 
the Spanish Ministry of Economy, Industry, and Competitiveness (MINECO) 
and the European Regional Development Fund FEDER Grant No. FIS2015-67161-P (MINECO/FEDER, UE).


\bibliography{biblioteka.bib}

\end{document}